\documentclass{elsart}
\usepackage{graphicx}
\usepackage{amsmath}
\usepackage{amssymb}
\usepackage[german,english]{babel}
\begin{document}
\sloppy
\begin{frontmatter}
\frenchspacing
\title{Determination of Interaction Potentials in Freeway Traffic from Steady-State Statistics}

\author[a,b,c,d]{Milan Krbalek}
\author[d]{Dirk Helbing}
 
\address[a]{Institute of Physics, Czech Academy of Science,
    Cukrovarnick\'a 10, 162 53 Prague, Czech Republic}
\address[b]{Faculty of Nuclear Sciences and Physical Engineering,
      Trojanova 13, 120 00 Prague, Czech Republic}
\address[c]{Max Planck Institute for the Physics of Complex Systems,
      Noethnitzer Str. 38, 01187 Dresden, Germany}
\address[d]{Institute for Economics and Traffic, Dresden University of
    Technology, Andreas-Schubert Str. 23, 01062 Dresden, Germany}

\begin{abstract}Many-particle simulations of vehicle interactions have been quite
successful in the qualitative reproduction of observed traffic
patterns. However, the assumed interactions could not be measured,
as human interactions are hard to quantify compared to
interactions in physical and chemical systems. We show that progress can be made
by generalizing a method from equilibrium statistical physics we learned from random matrix theory.
It allows one to determine the interaction potential via distributions of the netto distances $s$ of
vehicles. Assuming power-law interactions, we find that driver behavior can be approximated by a
forwardly directed $1/s$ potential in congested
traffic, while interactions in free traffic are characterized by an exponent of $\alpha \approx 4$. This is
relevant for traffic simulations and the assessment of telematic systems.
\end{abstract}
\begin{keyword}
Freeway traffic, power law interaction potential, random matrix theory, Dyson's gas, adaptive driver behavior,
optimal velocity model, approximate Hamiltonian, distance distribution, velocity distribution\\[3mm]
\end{keyword}
\end{frontmatter}

\section{Introduction}

In recent years, statistical physics and nonlinear dynamics have been very successful in
modeling, simulating, and understanding empirically observed instability and pattern 
formation phenomena in freeway traffic \cite{general,Review,TGF01,phase,pattern,gas}.
On the one hand,
vehicle traffic flow can be well approximated as a particular one-dimensional particle gas 
with Boltzmann statistics \cite{gas,Review}. On the other hand, 
the motion of a vehicle $i$ of mass $m$ and length $l_i$ at location $r_i(t)$
with speed $v_i(t) = dr_i/dt$, maximum velocity $v_0$,
adaptation time $\tau$, and fluctuation force $\xi_i(t)$ can be
reflected by an acceleration equation of the form
\begin{equation}
 m\frac{dv_i}{dt} = m \frac{v_0 - v_i}{\tau} + f(s_i) + \xi_i(t) \, .
\label{OV} 
\end{equation}
$f(s_i)$ is a negative and repulsive force, which depends on the netto 
distance (clearance) $s_i = (r_{i+1} - r_i - l_i)$ between two successive vehicles. 
With $W_i(s) = v_0 + \tau f(s)/m$, Eq. (\ref{OV}) corresponds to the optimal velocity
model $dv_i/dt = [W_i(s_i) - v_i]/\tau$ \cite{Bando} with an additional noise term $\xi_i$. 
Due to a lack of good single-vehicle data and suitable evaluation methods, 
the form of the interaction potential $U(s)$ with $f(s) = - dU(s)/dr 
= - dU(s)/ds$ has been subject to speculation. A direct evaluation by averaging $v_i$-over-$s_i$ data
suffers from the wide distribution of netto distances and causes problems of interpretation
\cite{Review,twophase}. Exploiting the source of this problem, we suggest a {\em statistical} approach
to determine the potential and present first results regarding its form,
based on a comparison of single vehicle data with the netto distance
distributions obtained from rigorous solutions for a particle gas in equilibrium.
For a justification of this approach see Ref.~\cite{fpg}.
\par
Normally, the interaction forces or potentials between particles
governing each other's motion, are not directly
measurable. The great success of scattering theory was to
determine interaction potentials from  statistical distributions
of particles scattered at some ``target'' composed of the material
under investigation \cite{scatter}. Learning about human
interactions requires a somewhat different approach from statistical physics, which
can be learned from books on random matrix theory \cite{Mehta}. This method was successfully 
applied to the statistical description of the time gaps $T_i$ between the arrival times of buses in
some Mexican cities \cite{Seba}, where the potential
$U$ was found to be a logarithmic function, i.e. $U(T_i)=-\ln (T_i).$ It was speculated that the
same potential would approximate the (time) headway
distribution of highway traffic \cite{Wagner}. In
contrast, our study will identify the shape of the spatial
interaction potential and reveal a different driver behavior in
free and congested traffic.

\section{Methodological Approach}

We have recently extended a method developed for classical many-particle
systems exposed to a ``thermal bath'' of a given temperature, i.e. to
random forces of a certain variance and statistics \cite{fpg}. The resulting
velocity  and netto distance distributions [see Eqs.~(\ref{Vel}) and (\ref{Headway})] allow
one to draw conclusions about the interaction potential $U,$ as
this determines their shapes. These distributions have been originally derived assuming a conservation
of momentum and energy, i.e. a transformation of potential energy 
into kinetic energy. 
However, by investigating a Fokker-Planck equation equivalent to Eq.~(\ref{OV}),
one can show that the same distributions are steady-state solutions for driven many-particle
systems with forwardly directed rather than symmetrical potentials \cite{fpg}, if the system
is large enough and if the average velocity $V$ and variance $\theta$ are constant.
For $dV/dt \approx 0$, $d\theta/dt \approx 0$, and $N \gg 1$, 
fluctuations in the system become negligible, and the total energy 
is only slightly fluctuating. The effects of the driving force $m v_0/\tau$ and the dissipative force
$- m v_i/\tau$ balance each other in a statistical sense \cite{fpg}. However, $dV/dt \approx 0$ requires
linear stability of the system (\ref{OV}) of coupled differential equations, i.e.
$dW(s)/ds < 1/(2\tau)$ \cite{Bando}. Therefore, we have to exclude the 
stop-and-go regime between 20 and 40 vehicles per kilometer and lane from our investigation
\cite{Review}. In summary, the data analysis must be carried out  for ensembles of vehicles
in a stationary traffic state with a well-defined
velocity variance (generalized ``temperature'') Var$(v_i) = (\sigma V)^2$
in a reference frame moving with a constant velocity, the average vehicle velocity $V$.
That is, one has to restrict to small density intervals, otherwise one will mix
up systems of different generalized temperatures and different average velocities. 
Such a careful analysis is presented here for the first time for single vehicle data of
the Dutch two-lane freeway A9,
which was confirmed by another analysis for Czech highway data (not shown).
This has led to new and surprising insights regarding the shape of the interaction potential
and its dependence on the traffic state.
We will focus on forwardly directed power-law potentials
\begin{equation}
U(s) \propto \left\{ 
\begin{array}{cc}
s^{-\alpha} & \mbox{for } s > 0 \\
0 & \mbox{otherwise} \, ,
\end{array}
\right.
\label{pote}
\end{equation}
where $\alpha > 0$ is a fit parameter and $s$ the netto distance (clearance) between
two successive cars on a (ring) road of length $L$. Similar relations 
have been suggested in some car-following models \cite{carfollow,IDM} and describe the
repulsive tendency of drivers to keep a safe distance to the respective car ahead. 
The exponent determines the characteristic dynamic and stationary behavior. 

\section{Short-ranged Dyson's gas with power-law potential}

We will now investigate the statistical gas of $N$ point-like
identical particles on a ring of scaled length $N$ interacting via the potential energy \cite{print}
\begin{equation}
{\mathcal{U}}(x_1,\ldots,x_N)= C \sum_{i=1}^{N} U(x_{i+1}-x_i) \, .
\label{Potential}
\end{equation}
Herein, $(x_1,\ldots,x_N)$ is the
vector of scaled particle positions 
\begin{equation}
x_i = (r_i - \sum_{j=1}^{i-1} l_j) \frac{N}{L}  = \rho (r_i- \sum_{j=1}^{i-1} l_j) \, ,
\end{equation} 
where $\rho$ denotes the global vehicle density. For
convenience, we will use a periodic index, i.e. $x_{i+N}=x_i + N.$
Let this so-called short-ranged power-law Dyson's gas (SRDG) be
exposed to a temperature reservoir with Boltzmann statistics and generalized temperature
$\sigma^2 > 0.$ The corresponding dimensionless Hamiltonian reads
\begin{equation}
{\mathcal{H}}=\frac{1}{2} \sum_{i=1}^N (u_i-\langle u \rangle )^2 +  
\frac{C}{mV^2} \sum_{i=1}^{N} U(x_{i+1}-x_i) 
\end{equation}
with the additional condition $\sum_{i=1}^{N}|x_{i+1}-x_i|=N$, 
where $u_i = v_i \langle u \rangle/V$ is the
scaled velocity of the $i$th particle and $\langle u \rangle = 1$ the scaled
average velocity. Note that 
all the quantities used here
are dimensionless. The probability of finding the system
in the phase-space element $\Omega \equiv (x_1,\ldots,x_N,u_1,\dots,u_N)$ is
\begin{eqnarray}
P_*(\Omega)&=&{\mathcal{N}}e^{-{\mathcal{H}}/(2\sigma^2)}={\mathcal{N}}\prod_{i=1}^N
e^{- (u_i-\langle u \rangle)^2 / (2\sigma^2)} \nonumber \\
&\times& \prod_{i=1}^{N}e^{-\beta U(x_{i+1}-x_i)} \,
\delta\left(N - \sum_{i=1}^{N}|x_{i+1}-x_i|\right) \, ,
\end{eqnarray}
where $\beta = C/(m \sigma^2 V^2)$ represents  the scaled generalized inverse temperature
and $\mathcal{N}$ the normalization factor, which can be
determined from the normalization condition $\int_{\Bbb{R}^{2N}}
P_*(\Omega)\,d\Omega=1$ \cite{print}. By integration over the $2N-1$ independent
variables, the probability of finding the $i$th particle with
velocity $u_i$ is obtained as
\begin{equation}
P'(u_i)=\frac{1}{\sqrt{2\pi}\sigma}\,
\exp\left[ -\frac{(u_i-\langle u \rangle)^2}{2\sigma^2} \right] \, .
\label{Vel}
\end{equation}
That is, the velocities are Gaussian-distributed with mean value 
\begin{equation}
\langle u \rangle \equiv
\int_{\Bbb{R}} u P'(u)\,du = 1
\end{equation} 
and variance 
\begin{equation}
\langle (u-\langle u \rangle)^2 \rangle
\equiv \int_{\Bbb{R}} (u-\langle u \rangle)^2 P'(u)\,du = \sigma^2 = \langle (v-V)^2 \rangle / V^2 \, .
\end{equation}
\par
For the above SRDG, one can also determine the equilibrium distribution of scaled
particle distances $z_i= (x_{i+1}-x_i) = \rho s_i$ \cite{Krbalek},
which gives us the expression for the netto distance (or clearance) distributions:
\begin{equation}
P_\beta^{(\alpha)} (z)=A\,e^{-\beta z^{-\alpha}}e^{-Bz} \, .
\label{Headway}
\end{equation}
Herein, $A=A(\alpha,\beta)$ and $B=B(\alpha,\beta)$ are
constants determined via the normalization conditions 
\begin{eqnarray}
\int_0^\infty P_\beta^{(\alpha)} (z)\,dz=1 \, , \quad
\langle z \rangle \equiv \int_0^\infty z\,P_\beta^{(\alpha)}
(z)\,dz=1 \, . 
\end{eqnarray}
These equations can be analytically solved only for particular
potentials, including power-law relations. In the following,
we need two special variants of the SRDG. In the case $\alpha=1$, we find
\begin{equation}
\int_0^\infty P_\beta^{(1)} (z)\,dz
=2A\sqrt{\frac{\beta}{B}} {\mathcal{K}}_1(2\sqrt{\beta B})
\end{equation} 
and
\begin{equation}
\int_0^\infty z \,P_\beta^{(1)} (z)\,dz =2A\frac{\beta}{B} {\mathcal{K}}_2(2\sqrt{\beta B}) \, , 
\end{equation}
where ${\mathcal{K}}_\lambda$ is the Mac-Donald's function
(modified Bessel's function of the second kind) of order $\lambda
\in \Bbb{R}.$ Based on these equations, one can exactly determine the normalization
constants $A$ and $B$. 
For the case $\alpha=4$, the numerically determined values
of the normalization constants $A$ and $B$ are displayed in 
Table~\ref{Tab1}. For $\beta > 2$, we could find the 
approximate relations
\begin{equation}
A \approx \exp\left(\alpha\,\beta - 0.1490\,\alpha^2 +
1.3689\,\alpha + 0.2271\right)
\end{equation} 
and
\begin{equation} 
B \approx \alpha\,\beta +0.4593\,\alpha + 0.9481 \, .
\end{equation}

\section{Data analysis} 

\begin{figure}[htbp]
\begin{center}
\hspace*{-2mm}\includegraphics[width=7.5cm, angle=0]{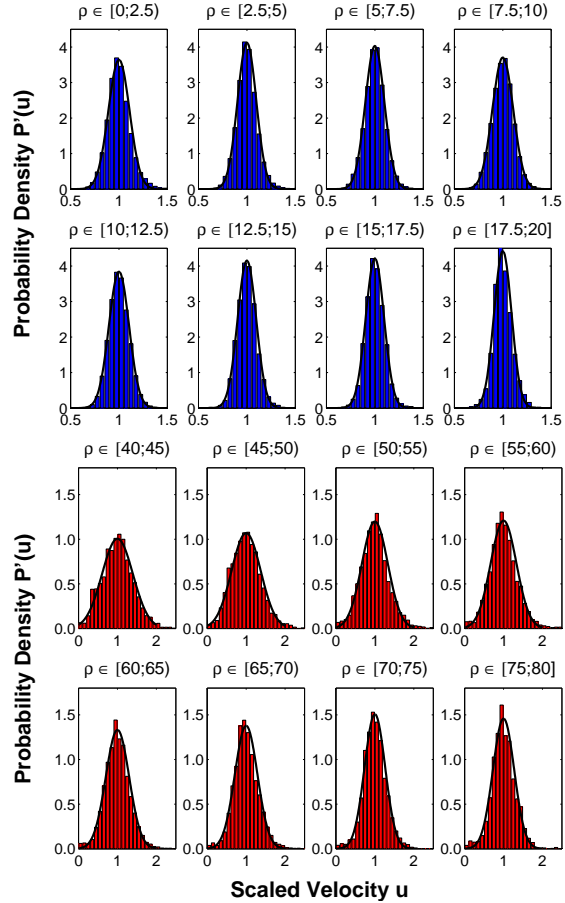}
\end{center}
\caption{Scaled velocity distributions for eight density regimes in free traffic (above)
and eight density intervals in congested traffic (below).
The bar diagrams correspond to the scaled empirical velocity distributions, while
the solid curves correspond to the theoretically predicted
Gaussian distributions. Note that the mean speeds are always scaled to one, while
the variances are given by the scaled empirical values $\sigma^2$ 
(see Table \ref{Tab2}).\label{veldist}}
\end{figure}

\begin{figure}[htbp]
\begin{center}
\hspace*{-2mm}\includegraphics[width=7.5cm, angle=0]{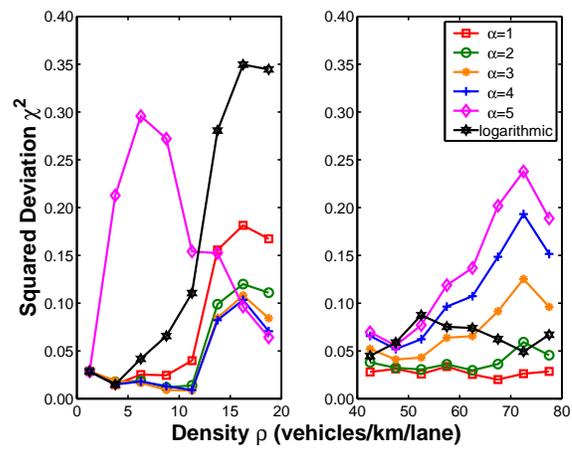}
\end{center}
\caption{Sum of squared deviations between the empirical and the 
theoretical netto distance distributions for various fit parameters $\alpha$.
The best fit is $\alpha = 4$ throughout the free traffic regime (left) and $\alpha = 1$ 
throughout the congested regime (right).\label{leastsq}}
\end{figure}

\begin{figure}[hptb]
\begin{center}
\hspace*{-2mm}\includegraphics[width=7.5cm, angle=0]{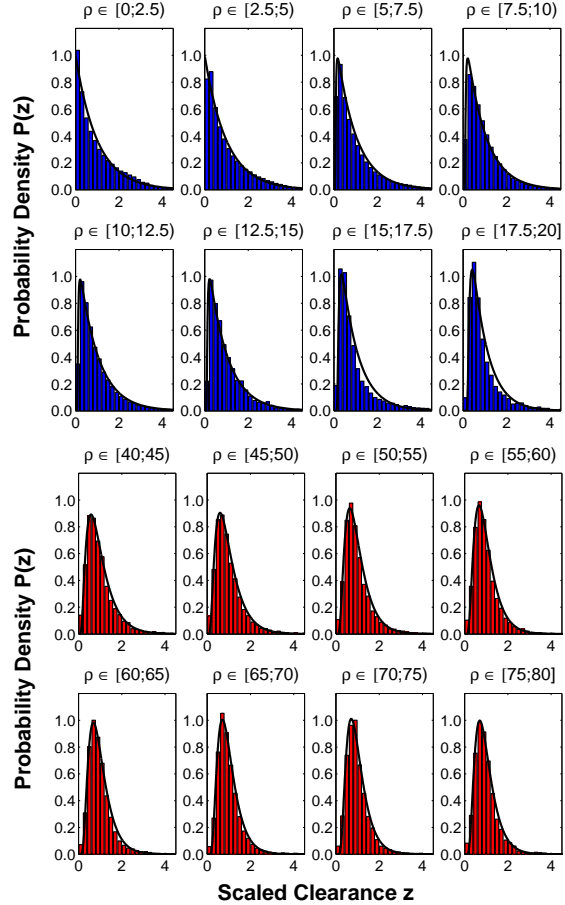}
\end{center}
\caption{Distributions of scaled netto distances (clearances) $z = \rho s$
for eight density intervals in free
traffic (above) and eight density regimes in congested traffic (below).
The bar diagrams represent the scaled empirical distributions,
while the solid curves correspond to the normalized theoretical distributions
with the empirically determined values $\beta$ listed in Table~\ref{Tab2} and parameter 
$\alpha = 4$ for free traffic, but $\alpha = 1$ for congested traffic.\label{distdist}}
\end{figure}

We have separately analyzed eight small density intervals 
in the free low-density regime $\le 20$ veh./km and eight density intervals
in the congested traffic regime $\ge 40$ veh./km. After determination of the respective values of 
$\beta$ and $\sigma$ from the single vehicle data, we have obtained the fit parameter
$\alpha$ by a least square method, i.e. minimization of the error function $\chi^2$ measuring the
deviation between the theoretical and empirical clearance distributions.
The predicted Gaussian distributions reproduce the empirical velocity distributions 
very well (see Fig.~\ref{veldist}), which is
also supported by other empirical and numerical studies \cite{gas}.
The best fit of the netto distance distributions is obtained for the integer parameter $\alpha = 1$
in congested traffic, which is, for example, compatible with the
intelligent driver model (IDM) and perception-based models \cite{IDM}. Throughout the free
traffic regime, we find a good agreement with $\alpha =  4$, corresponding
to weak interactions \cite{print} (see Figs.~\ref{leastsq}, \ref{distdist}).
This is  not only a strong support of
studies questioning a uniform behavior of drivers in all traffic 
regimes \cite{TGF01,twophase}. It also offers
an interpretation of the mysterious fractional distance-scaling exponent $\alpha+1 \approx
2.8$ in  classical follow-the-leader models \cite{carfollow}, which interpolate between the driver behavior
in the free and congested regimes. 

\section{Summary and Discussion}

We have found that it is successful to generalize thermal-equilibrium
properties of a short-ranged power-law Dyson's gas exposed to
a heat reservoir with the scaled generalized inverse temperature $\beta$ to steady-state
vehicle traffic, where $\beta$ is an increasing function of the traffic density $\rho.$ In the regime of
congested traffic, this dependency is simply linear: $\beta(\rho) = 0.0261\rho - 0.4785$.
The presented results show that the shape of the interaction potential of vehicles
can be approximated by formula (\ref{pote}) with $\alpha=4$ for free traffic and
$\alpha=1$ for congested traffic.  The theoretical predictions are further supported by the Gaussian
distribution of the vehicle velocities in all investigated density regimes. 
\par
As it is a hard task to derive analytical relations for the clearance distribution (including its
normalization constants $A$ and $B$), we could demonstrate the determination of
vehicle interaction potentials only for simple functional relations. Future advances with this method
will hopefully allow to determine velocity-dependent interaction forces or functions with turning points, 
which are desireable to reproduce the dynamical behavior in the unstable traffic regime realistically 
as well \cite{turning}. Determining interaction potentials in freeway traffic
does not only contribute to the challenging problem of how to fit
time headway or distance distributions of vehicles
\cite{twophase,headways,Review}. It also advances 
the quantitative understanding of human behavior. Normally, it is
difficult to identify and measure the relevant behavioral
variables, and realistic models contain a large number of
parameters. Here, progress has been made by powerful
methods from statistical physics. Nevertheless, the resulting
interaction model is not just a physical model. The new
picture of two different regimes with $\alpha = 1$ and $\alpha = 4$ points to adaptive
driver behavior to congested and free traffic conditions. The specification of the interaction
potential is essential for realistic traffic simulations, which
are required for the design of efficient and reliable traffic
optimization measures such as intelligent on-ramp controls, driver
assistance systems, lane-changing assistants, etc.

\subsection*{Acknowledgements} 

This work was supported by the Czech Academy of Sciences under Grant No. A1048101 and by the
``Foundation for Theoretical Physics'' in Slemeno, Czech Republic. 
The authors would also like to thank
Henk Taale and the {\em Dutch Ministry of Transport, Public Works
and Water Management} as well as P{e}tr \v{S}eba, Ji\v r\'{\i} M\'{\i}\v
sek, Petr Persa\v n, and the {\em Road and Motorway Directorate of
the Czech Republic} in Brno for providing the single-vehicle
induction-loop-detector data.

\clearpage

\begin{table}[htbp]
\begin{tabular}{|c|c|c||c|c|c|}\hline
 $\beta$ & $A$ & $B$ & $\beta$ & $A$ & $B$ \\
\hline 
0&1&1&                                 0.001&1.6558&1.2644\\
0.00001&1.158&1.0761 &   0.005&2.2673&1.4405\\ 
0.00005&1.246&1.1123 &   0.01&2.783&1.5593\\
0.0001&1.3036&1.136  &    0.05&6.1623&2.0447\\
0.0005&1.5122&1.2151 &   0.1&11.1941&2.4296\\
 \hline
\end{tabular}
\caption[]{Numerically determined values of the normalization
constants $A$ and $B$ for $\alpha =4$. The values depend on $\beta$, and
therefore, also on the density $\rho$ (see Table~\ref{Tab2}).\label{Tab1}}
\end{table}

\begin{table}[htbp]
\begin{tabular}{|c|c|c||c|c|c|}\hline
 $\rho$ & $\sigma$ & $\beta$ &  $\rho$ & $\sigma$ & $\beta$ \\
 (veh./km) &  $(10^{-1})$ & $(10^{-5})$ & (veh./km) & $(10^{-1})$ & $(10^{-5})$ \\
\hline
$[0;2.5)$ & 1.095 & $\approx$ 0  & $[2.5;5)$ & 0.967  & $\approx$ 0 \\ 
$[5;7.5)$ & 0.989  & 3.998 & $[7.5;10)$ & 1.076 & 7.331 \\
$[10;12.5)$ & 1.038 & 11.86 & $[12.5;15)$ & 0.960 & 16.36 \\
$[15;17.5)$ & 0.946 & 152.9 & $[17.5;20]$  & 0.902 & 442.4 \\
 \hline
$[40;45)$  & 3.967 & 634.8 & $[45;50)$ & 3.731 & 711.8  \\
$[50;55)$ & 3.329  & 915.9 & $[55;60)$ & 3.300 & 1046 \\
$[60;65)$ & 2.996 & 1186 & $[65;70)$ & 2.895 & 1323 \\
$[70;75)$ & 2.647 & 1358 & $[75;80]$ & 2.735 & 1279 \\
\hline
\end{tabular}
\caption[]{Empirical values of the scaled velocity variation
$\sigma$ and the scaled generalized inverse temperature
$\beta$ for 16 density intervals obtained from single-vehicle data of
the Dutch two-lane freeway A9, neglecting clearances in front of long vehicles
with $l_i > 7$~m.\label{Tab2}}
\end{table}

\end{document}